\documentclass[12pt,preprint]{aastex}
\def\la{\mathrel{\hbox{\rlap{\hbox{\lower4pt\hbox{$\sim$}}}\hbox{$<$}}}}

\begin{document}

% ******************************************************************
\title{Constraining $H_0$ from Sunyaev-Zel'dovich effect, Galaxy Clusters X-ray data, and
Baryon Oscillations}
\author{J. V. Cunha}
\affil{Departamento de Astronomia, Universidade de S\~ao Paulo, USP,
\\ 05508-900 S\~ao Paulo, SP, Brazil}
\email{cunhajv@astro.iag.usp.br}
\author{L. Marassi}
\affil{Departamento de F\'{\i}sica, Universidade Federal do Rio
Grande do Norte, \\ C.P. 1641, Natal, RN, 59072-970, Brazil}
\email{lucio@dfte.ufrn.br}
\author{J. A. S. Lima}
\affil{Departamento de Astronomia, Universidade de S\~ao Paulo, USP,
\\ 05508-900 S\~ao Paulo, SP, Brazil}
\email{limajas@astro.iag.usp.br}

% ******************************************************************
%    Abstract
% ******************************************************************
\begin{abstract}

Estimates of $H_0$ from Sunyaev-Zel'dovich effect (SZE) and X-ray surface brightness of galaxy clusters depends on the underlying
cosmology. In the current $\Lambda$CDM flat cosmology,  a possible
technique to broke the degenerescency on the mass density
parameter ($\Omega_{m}$) is to apply a joint analysis involving
the baryon acoustic oscillations (BAO). By adopting this technique
to the ($H_0, \Omega_m$) parameter space, we obtain new
constraints on the Hubble constant $H_0$ from BAO signature as
given by the Sloan Digital Sky Survey (SDSS) catalog. Our analysis
based on the SZE/X-ray data for a sample of 25 clusters yields
$H_0= 74^{+4}_{-3.5}$ km s$^{-1}$ Mpc$^{-1}$ ($1\sigma$,
neglecting systematic uncertainties).  This result is in good
agreement with independent studies from the {\it{Hubble Space
Telescope}} key project and the recent estimates of WMAP, thereby
suggesting that the combination of these three independent
phenomena provides an interesting method to constrain the Hubble
constant.
\end{abstract}
% ******************************************************************

% ******************************************************************
% 1. INTRODUCTION
% ******************************************************************
\section{INTRODUCTION}
\label{sec:intro}
Galaxy clusters are one of the most impressive evolving structures
from an earlier stage of the Universe. Usually, they congregate
thousands of galaxies and are endowed with a hot gas (in the intra
cluster medium), emitting X-rays primarily through thermal
bremsstrahlung. Several studies in the last decade have suggested
that the combination of data from different physical processes in
galaxy clusters provides a natural method for estimating some
cosmological parameters (Bartlett \& Silk 1994, Rephaeli 1995,
Kobayashy {\it{et al.}} 1996, Reese {\it{et al.}} 2002; Barttlet
2004; De Filippis {\it{et al.}} 2005; Bonamente {\it{et al.}}
2006). The ultimate goal in the near future is to shed some light
on the nature of the dark energy.

An important phenomena occurring in clusters is the
Sunyaev-Zel'dovich Effect (SZE), a small distortion of the Cosmic
Microwave Background (CMB) spectrum provoked by the inverse
Compton scattering of the CMB photons passing through a population
of hot electrons (Sunyaev \& Zel'dovich 1972). Since the SZE is
insensitive to the redshift of galaxy clusters, it provides a very
convenient tool for studies at intermediate redshifts where the
abundance of clusters depends strongly on the underlying cosmology
(the unique redshift dependence appear in the total SZE flux due
to the apparent size of the cluster). Another fundamental process
is the X-ray emission from the hot electrons in the intracluster
medium. When the X-ray surface brightness is combined with the SZE
temperature decrement in the CMB spectrum, the angular diameter
distance of galaxy clusters is readily obtained.

The possibility to estimate the galaxy cluster distances trough
SZ/X-ray technique was suggested long ago by many authors (Silk \&
White 1978; Birkinshaw 1979; Cavaliere {\it{et al.}} 1979), but
only recently it has been applied for a fairly large number of
clusters (for reviews, see Birkinshaw 1999; Carlstrom, Hoder \&
Reese 2002). Such a method is based on the different dependence of
the cluster electron density ($n_e$) and the temperature $T_e$ of
the SZE ($\propto n_eT_e$) and X-ray bremsstrahlung ($\propto
n_e^2T_e^{1/2}$). Combining both measurements it is possible to
estimate the angular diameter distance and infer the value of the
Hubble constant whether the cosmology is fixed. The main advantage
of this method for estimating $H_0$ is that it does not rely on
extragalactic distance ladder being fully independent of any local
calibrator. A basic disadvantage rests on the difficulty of
modeling the cluster gas which causes great systematic
uncertainties in its determination. In particular, this means that
systematic effects on $H_0$ are quite different from the ones
presented by other methods, like the traditional distance ladder
or gravitational lensing (Reese {\it{et al.}} 2002; Jones {\it{et
al.}} 2005; De Filippis {\it{et al.}} 2005).

In order to estimate the distance to the cluster from its X-ray
spectroscopy, one needs to add some complementary assumptions about
its geometry. The importance of the intrinsic geometry of the
cluster has been emphasized by many authors (Fox \& Pen 2002; Jing
\& Suto 2002; Plionis {\it{et al.}} 2006; Sereno {\it{et al.}}
2006). The standard spherical geometry has been severely questioned,
since Chandra and XMM-Newton observations have shown that clusters
usually exhibit an elliptical surface brightness. In a point of
fact, the cluster shape estimation problem is a difficult matter
since many clusters do not appear in radio, X-ray, or optical.
Another source of difficulty is related to the error bars. Assuming
that the clusters have an axisymmetric form, different authors
introduced a roughly random uncertainty in $H_0$ between $15\% -
30$\% (Hughes \& Birkinshaw 1998; Sulkanen 1999; Reese {\it{et al.}}
2002; Jones {\it{et al.}} 2005). The assumed cluster shape also
affects considerably the SZE/X-ray distances, and, therefore, the
Hubble constant estimates.

Fox and Pen (2002) estimate the Hubble constant by assuming
triaxial clusters and measuring the distance to artificial
observations corrected for asphericity. De Filippis and
collaborators (2005) showed that the spherical hypothesis is
strongly rejected for most members of the sample studied. By
taking into account such an effect for two samples, a better
agreement with the cosmic concordance model ($\Omega_m=0.3$,
$\Omega_{\Lambda}=0.7$) was obtained. Triaxial clusters  may also
be useful for reconciling the observed discrepancies in the total
mass of clusters as computed with lensing and X-ray measurements
(in this connection see Bonamente {\it{et al.}} 2006).

The determination of $H_0$ has a practical and theoretical
importance to many astrophysical properties of galaxies and quasars,
and several cosmological calculations, like the age of the Universe,
its size and energy density, primordial nucleosynthesis, and others
(Freedman 2000; Peacock 1999). Spergel {\it{et al.}} (2006) have
shown that CMB studies can not supply strong constraints on the
value of $H_0$ on their own. This problem occurs due to the
degenerescency on the parameter space (Tegmark {\it{et al}} 2004),
and may be circumvented  only  by using independent measurements of
$H_0$ (Hu 2005).

On the other hand, according to cold dark matter (CDM) picture of
structure formation, large-scale fluctuations have grown since
$z\sim1000$ by gravitational instability. The cosmological
perturbations excite sound waves in the relativistic plasma,
producing the acoustic peaks in the early universe. Eisenstein
{\it{et al.}} (2005) presented the large scale correlation
function from the Sloan Digital Sky Survey (SDSS) showing clear
evidence for the baryon acoustic peak at $100 h^{-1}$ Mpc scale,
which is in excellent agreement with the WMAP prediction from the
CMB data. The Baryon Acoustic Oscillations (BAO) method is
\emph{independent of the Hubble constant} $H_0$ which means that
we can use BAO signature to break the degenerescency of the mass
parameter $\Omega_m$. Hence, combining SZE/X-ray method to obtain
${\cal{D}}_A$ with BAO  it is possible to improve the limits over
$H_0$ (for recent applications of BAO see, Lima {\it{et al.}}
2006).

In this letter, by assuming that the clusters are ellipsoids with
one axis parallel to the line of sight, we derive new constraints on
the Hubble constant $H_0$. By considering the sample of 25 triaxial
clusters given by De Filippis {\it{et al.}} (2005), we perform a
joint analysis combining the data from SZE and X-ray surface
brightness with the recent SDSS measurements of the baryon acoustic
peak (Eisenstein {\it{et al.}} 2005).

\section{Basic equations and Sample}

Let us now consider that the Universe is described by a flat
Friedmann-Robertson-Walker (FRW)  geometry driven by cold dark
matter plus a cosmological constant. In this case, we have only
two free parameters ($H_0,\Omega_{m}$) and the angular diameter
distance, ${\cal{D}}_A$ reads (Lima {\it{et al.}} 2003, Alcaniz
2004, De Filippis {\it{et al.}} 2005)
\begin{equation}
{\cal{D}}_A(z;h,\Omega_m) = \frac{3000h^{-1}}{(1 +
z)}\int_{o}^{z}\frac{dz'}{{\cal{H}}(z';\Omega_m)} \quad \mbox{Mpc},
\label{eq1}
\end{equation}
where $h=H_0/100$ km s$^{-1}$ Mpc$^{-1}$ and the dimensionless
function ${\cal{H}}(z';\Omega_m)$ is given by
\begin{equation}
{\cal{H}} = \left[\Omega_m(1 + z')^{3} + (1 -\Omega_m)\right]^{1/2}.
\label{eq2}
\end{equation}

Following De Filippis {\it{et al.}} (2005), a general triaxial
morphology it will adopted here. In this case, the intra cluster
medium is described by an isothermal triaxial $\beta$-model
distribution and the SZE decrement reads

\begin{eqnarray}
\label{eq:sze3} \Delta T_{SZ} &\equiv &T_{0} f(\nu, T_{\rm e})
\frac{ \sigma_{\rm T} k_{\rm B} T_{\rm e}}{m_{\rm e} c^2}n_{e0}
\sqrt{\pi}\nonumber \\ &\times&  \frac{\cal{D}_{ A}\theta_{\rm
c,proj}}{b^{3/4}}\sqrt{\frac{e_1 e_2}{e_{\rm proj}}}g(\beta),
\end{eqnarray}
where  $T_{0}=2.728 K$ is the CMB temperature, $T_e$ is the gas
temperature, $\sigma_T$ is the Thompson cross section, the factor
$f(\nu, T_e)$ accounts for frequency shift and relativistic
corrections, $n_{eo}$ is the central number density of the cluster
gas, $b$ is a function of the cluster shape and orientation,
$e_{\rm proj}$ is the axial ratio of the major to the minor axes
of the observed projected isophotes, $\theta_{c,\rm proj}$ is the
projection on the plane of the sky of the intrinsic angular core
radius, and $g(\beta)={\Gamma(3\beta-1/2)}/{\Gamma(3 \beta)}$
($\Gamma$ denotes the Gamma function).
% ******************************************************************
%   Figure 1: Angular Diameter Distance X Redshift
% ******************************************************************
\begin{figure}[p] % [t]
   \epsscale{1.1}
   \plotone{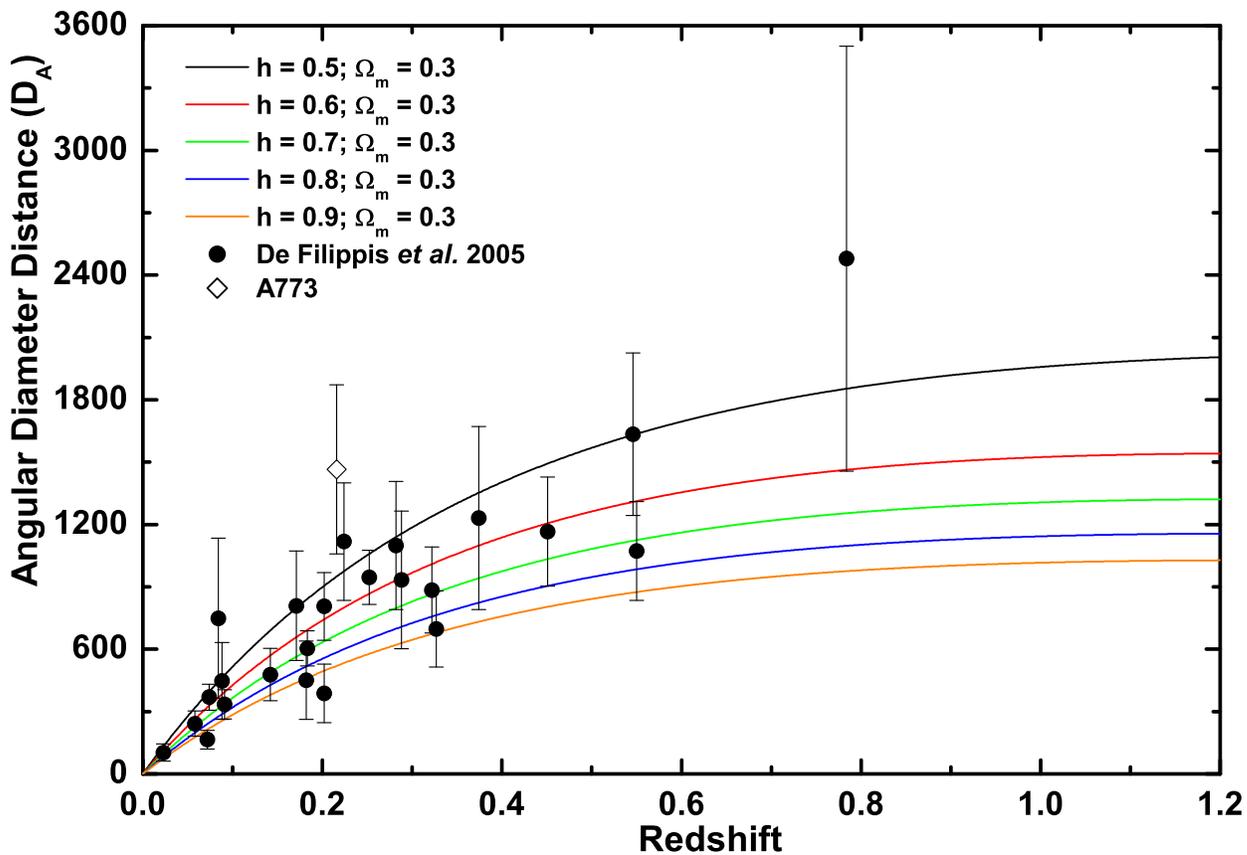}
   \caption{Angular diameter distance as a function of
redshift for $\Omega_{m}=0.3$ and some selected values of the $h$
parameter.  The data points correspond to the the SZE/X-ray
distances for 25 clusters from De Filippis {\it{et al.}} (2005).
The open diamond indicates the Abell 773 outlier cluster, which
has been excluded from our statistical analysis (see section 3).}
\label{Fig1}
\end{figure}

Similarly,  the X-ray central surface brightness $S_{X0}$ can be
written as
\begin{equation}
\label{eq:sxb2} S_{X0} \equiv \frac{ \Lambda_{eH}\ \mu_e/\mu_H}{4
\sqrt{\pi} (1+z)^4}\frac{n_{e0}^2 {\cal D_{A}}\theta_{\rm
c,proj}}{b^{3/4}}\sqrt{\frac{e_1 e_2}{e_{\rm proj}}}g(\beta),
\end{equation}
where $z$ is the redshift of the cluster,  $\Lambda_{eH}$ is the
X-ray cooling function of the ICM in the cluster rest frame and
$\mu$ is the molecular weight ($\mu_i\equiv \rho/n_im_p$).

De Filippis and collaborators (2005) studied and corrected the
${\cal{D}}_A$ measurements for 25 clusters, getting a better
agreement with the $\Lambda$CDM models. We used two samples studied
by them to investigate the bounds arising from SZE/X-ray
observations. One of the samples, compiled by Reese {\it{et al.}}
(2002), is a selection of 18 galaxy clusters distributed over the
redshift interval $0.14 < z < 0.8$. The other one, the sample of
Mason {\it{et al.}} (2001), has 7 clusters from the X-ray limited
flux sample of Ebeling {\it{et al.}} (1996). De Filippis {\it{et
al.}} (2005) show that the samples turn out slightly biased, with
strongly elongated clusters preferentially aligned along the line of
sight. Their results suggest that 15 clusters are in fact more
elongated along the line of sight, while the remaining 10 clusters
are compressed.

In Fig. 1, the galaxy cluster sample is plotted on a residual Hubble
diagram using a flat cosmic concordance model ($\Omega_m=0.3,
\Omega_{\Lambda}=0.7$). We see that the $A773$ cluster is the
largest outlier, and our statistical analysis confirms that its
inclusion leads to the highest $\chi^2$. For that reason we have
excluded this data point from the statistical analysis.

\section{Analysis and Results}

Now, let us perform a $\chi^2$ fit over the $h - \Omega_{m}$ plane.
In our analysis we use a maximum likelihood that can be determined
by a $\chi^2$ statistics,
\begin{equation}
\chi^2(z|\mathbf{p}) = \sum_i { ({\cal{D}}_A(z_i; \mathbf{p})-
{\cal{D}}_{Ao,i})^2 \over \sigma_{{\cal{D}}_{Ao,i}}^2},
\end{equation}
where ${\cal{D}}_{Ao,i}$ is the observational angular diameter
distance, $\sigma_{{\cal{D}}_{Ao,i}}$ is the uncertainty in the
individual distance and the pair, $\mathbf{p} \equiv (h,
\Omega_{m})$, is the complete set of parameters.

In what follows, we first  consider the SZE/X-ray distances
separately, and, further, we present a joint analysis including the
BAO signature from the SDSS catalog. Note that a specific flat
cosmology has not been fixed by hand in the analyzes below.

\subsection{Limits from SZE/X-ray}

We now consider the $24$  clusters (without the A773, see Fig.1),
which constitutes the SZE/X-ray data from De Filippis {\it{et al.}}
(2005). Our analysis indicated that any cosmological model could be
accepted by that sample until $3\sigma$ (with $2$ free parameters).
It also shows that using only the ellipsoidal cluster sample we
cannot constrain the energetic components of the cosmological model.
This happens basically because the error bars are large, mainly at
intermediate and high redshifts.
\begin{figure*}[p] % [t]
     \epsscale{1.1}
     \plotone{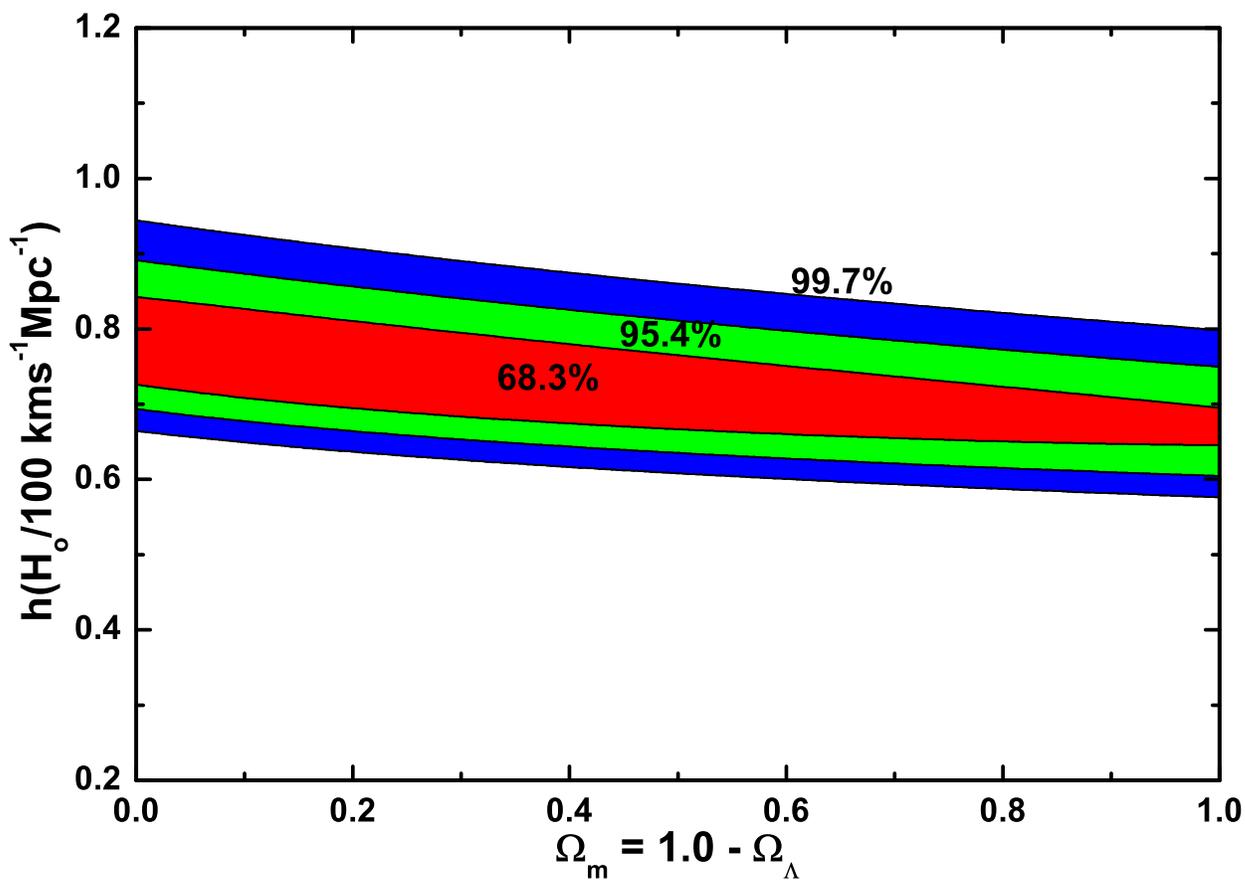}
     \caption{Confidence regions ($68.3$\%, $95.4$\% and
$99.7$\%) in the $(\Omega_{m}, h)$ plane provided by the SZE/X-ray
data from De Filippis {\it{et al.}} (2005). The best fit values are
$h = 0.75$ and $\Omega_{m} = 0.15$.} \label{Fig2}
\end{figure*}

In Fig. 2 we show the contours of constant likelihood (68.3\%,
95.4\% and 99.7\%) in the space parameter $h-\Omega_m$ for the
SZ/X-ray data discussed earlier.  Note that only a small range for
the $h$ parameter is allowed, ($0.64 \leq h \leq 0.85$), at
$1\sigma$ of confidence level. In particular, we found
$h=0.75^{+0.07}_{-0.07}$ and $\Omega_m = 0.15^{+0.57}_{-0.15}$
with $\chi^2_{min}=24.4$ at $68.3$\% c.l. for $1$ free parameter.
Naturally, such bounds on $h$ are reasonably dependent on the
cosmological model adopted. For example, if we fix
$\Omega_{m}=0.3$ we have $h=0.74$, for $\Omega_{m}=1.0$ we have
$h=0.67$, and both cases are permitted with high degree of
confidence. Clearly, we see that an additional cosmological test
(fixing $\Omega_m$) is necessary in order to break the
degenerescency on the $(\Omega_{m}, h)$ plane.

Systematic effects still need to be considered. The common errors
are: SZ $\pm 8$\%, X-ray $\pm 10$\%, radio halos $-3$\%, $5$\% for
Galactic N$_{H}$, $10$\% for isothermality, $2$\% kinetic SZ,
clumping causes $-20$\%, radio source confusion $\pm 12$\%, primary
beam $\pm 3$\% and $1$\% on the CMB. When we combine the errors in
quadrature, we find that the typical error are of $20$\% - $30$\%,
in agreement with others works (Mason {\it{et al.}} 2001; Reese
{\it{et al.}} 2002; Reese 2004).

\begin{figure*}[p] % [t]
     \epsscale{1.1}
     \plotone{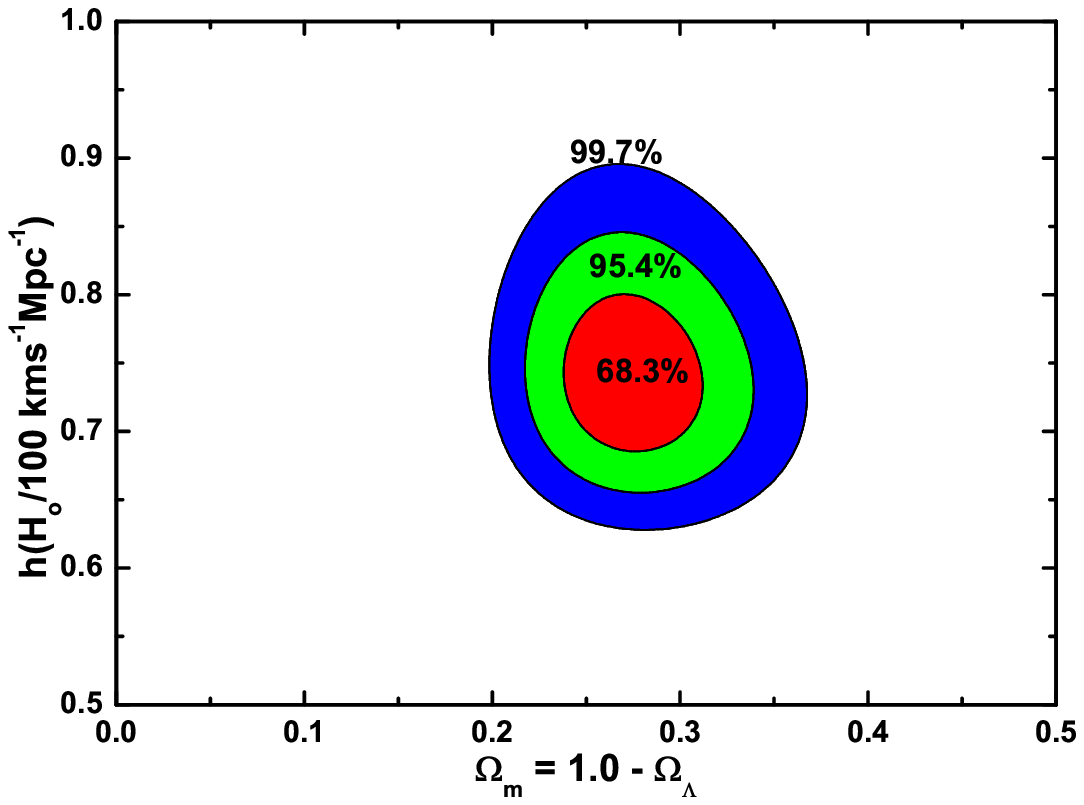}
     \caption{Contours in the $\Omega_m - h$ plane using
the SZE/X-ray and BAO joint analysis. The contours correspond to
$68.3$\%, $95.4$\% and $99.7$\% confidence levels. The best-fit
model converges to $h = 0.74$ and $\Omega_m=0.27$.} \label{Fig3}
\end{figure*}

\subsection{Joint Analysis for SZE/X-ray and BAO}

As remarked earlier, more stringent constraints on the space
parameter ($h, \Omega_m$) can be obtained  by combining the
SZE/X-ray with the BAO signature (Eisenstein {\it{et al.}} 2005).
The peak detected (from a sample of 46748 luminous red galaxies
selected from the SDSS Main Sample) is predicted to arise precisely
at the measured scale of 100 $h^{-1}$ Mpc. Basically, it happens due
to the baryon acoustic oscillations in the primordial baryon-photon
plasma prior to recombination. Let us now consider it as an
additional cosmological test over the ellipsoidal cluster sample.
Such a measurement is characterized by
\begin{eqnarray}
 {\cal{A}} \equiv {\Omega_{\rm{m}}^{1/2} \over
 {{\cal{H}}(z_{\rm{*}})}^{1/3}}\left[\frac{1}{z_{\rm{*}}}
 \Gamma(z_*)\right]^{2/3}  = 0.469 \pm 0.017, %\nonumber
\end{eqnarray}
where $z_{\rm{*}} = 0.35$ is the redshift at which the acoustic
scale has been measured, and $\Gamma(z_*)$ is the dimensionless
comoving distance to $z_*$.

Note that the above quantity is independent of the Hubble
constant, and, as such, the BAO signature alone constrains only
the $\Omega_m$ parameter. This property is very characteristic of
the BAO signature, thereby differentiating it from many others
classical cosmological tests, like the gas mass fraction (Lima
{\it{et al.}} 2003; Allen {\it{et al.}} 2004; Cunha {\it{et al.}}
2006), luminosity distance (Peebles \& Ratra 2003; Cunha {\it{et
al.}} 2002), or the age of the universe (Alcaniz {\it{et al.}}
2003; Cunha \& Santos 2004).

In  Fig. \ref{Fig3}, we show the confidence regions for the
SZE/X-ray cluster distance and BAO joint analysis. By comparing with
Fig. \ref{Fig2}, one may see how the BAO signature breaks the
degenerescency in the $(\Omega_{\rm{m}}, h)$ plane. As it appears,
the BAO test presents a striking orthogonality centered at
$\Omega_m= 0.27^{+0.03}_{-0.02}$ with respect to the angular
diameter distance data  as determined from SZE/X-ray processes. We
find $h= 0.738^{+0.042}_{-0.033}$ and $\chi^2_{min}=24.5$ at
$68.3$\% (c.l.) for $1$ free parameter. An important lesson here is
that the combination of SZE/X-ray with BAO provides an interesting
approach to constrain the Hubble constant.

\begin{figure*}[p] % [t]
     \epsscale{1.1}
     \plotone{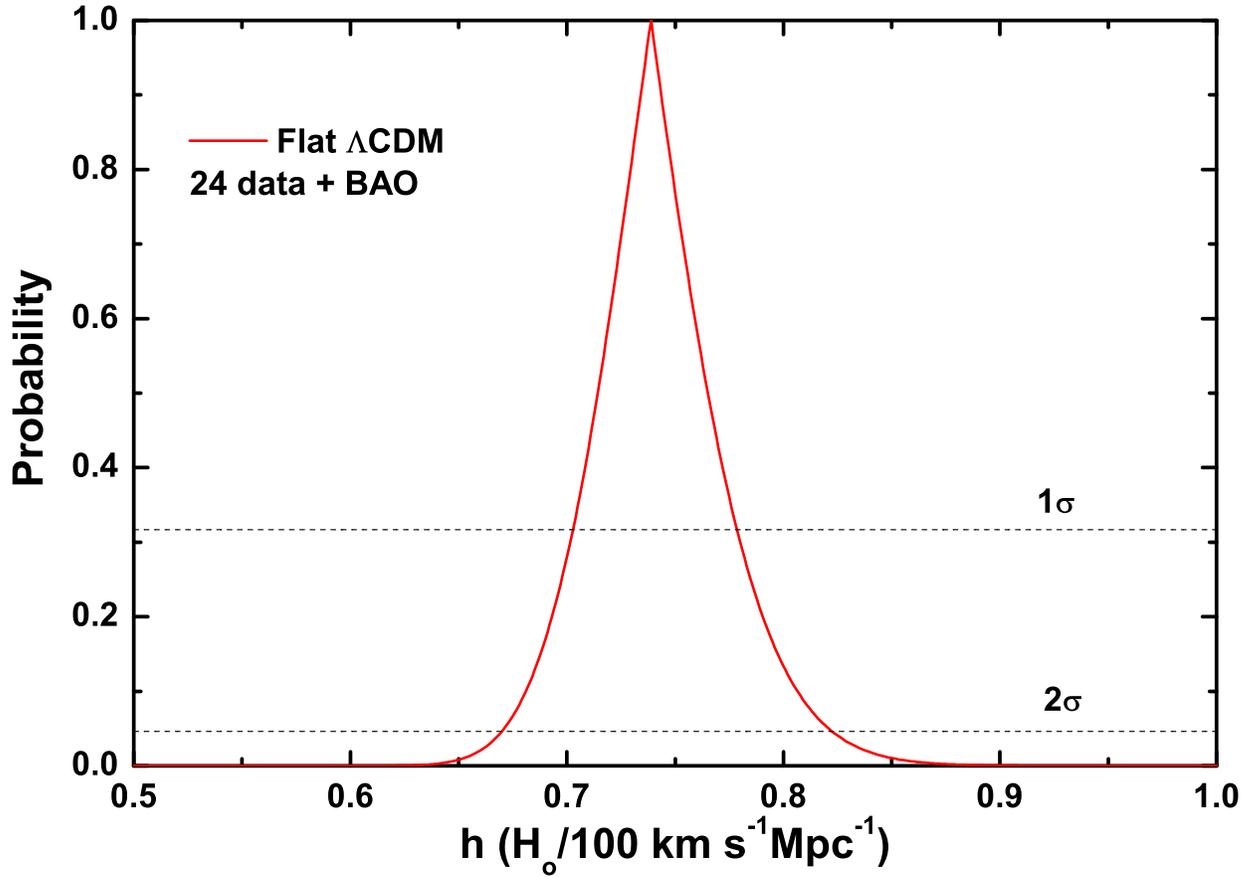}
     \caption{Likelihood function for the $h$ parameter in
a flat $\Lambda$CDM universe, from SZE/X-ray emission. The shadow
lines are cuts in the regions of $68.3$\% probability and $95.4$\%.
We see that the region permitted is well constrained and in
concordance with others studies (Freedman {\it{et al.}} 2001;
Spergel {\it{et al.}} 2006).} \label{Fig4}
\end{figure*}

In Fig. \ref{Fig4}, we have plotted the likelihood function for the
$h$ parameter in a flat $\Lambda$CDM universe for the SZE/X-ray +
BAO data set. The dotted lines are cuts in the regions of $68.3$\%
probability and $95.4$\%.

Our results are in line with some recent analyzes based on different
cosmological observations, like the one provided by the WMAP team
$h=0.73 \pm 0.03$ (Spergel {\it{et al.}} 2006), and the HST Project
$h=0.72 \pm 0.08$ (Freedman {\it{et al.}} 2001). Note, however, that
it does not agree with the recent determination, $h=0.62 \pm 0.013$
(random) $\pm 0.05$ (systematics), recently advocated by Sandage and
collaborators (2006). A result obtained with basis on Type Ia
Supernovae, calibrated with Cepheid variables in nearby galaxies
that hosted them.

At this point, it is interesting to compare our results with
others recent works in which the limits on $h$ were obtained by
fixing the cosmology ($\Omega_m=0.3$, $\Omega_{\Lambda}=0.7$,
cosmic concordance), and assuming spherical geometry. A
measurement using SZ effect was accomplished by Mason {\it{et
al.}} (2001), using 5 clusters, and gives
$h=0.66^{+0.14}_{-0.11}$; Reese and coauthors (2002), using 18
clusters, found $h=0.60\pm 0.04$, and in a posterior analysis
Reese (2004), with 41 clusters, obtains $h\approx 0.61\pm 0.03$;
Jones {\it{et al.}} (2005) derived $h=0.66^{+0.11}_{-0.10}$, using
a sample of 5 clusters free of any orientation bias. In a recent
paper, Bonamente {\it{et al.}} (2006), using 38 clusters, obtained
$h=0.769^{+0.039}_{-0.034}$. All these results, using SZ/X-ray
technique, presented a systematic uncertainty of $10\%$-$30\%$. In
Table 1, we summarize the estimates of $H_0$ from clusters in the
framework of $\Lambda$CDM models (the data in round brackets is
the number of clusters).

It is worth notice that the best-fit scenario derived here,
$\Omega_m= 0.27^{+0.03}_{-0.02}$ and  $h=
0.738^{+0.042}_{-0.033}$, corresponds to an accelerating Universe
with $q_0=-0.6$, a total evolutionary age of $t_o \simeq 10h^{-1}$
Gyr, and a transition redshift (from deceleration to acceleration)
$z_{t} \simeq 0.6$. At 95.4\% c.l. ($2\sigma$) the BAO+SZE/X-ray
analysis also provides $h= 0.74^{+0.08}_{-0.07}$. Hopefully,
future developments related to the physics of clusters may shed
some light on the nature of the dark energy (for reviews see
Peebles \& Ratra 2003; Padmanhaban 2003; Lima 2004).

\section{Conclusions}

Since the original work of Hubble, the precise determination of
the distance scale ($H_0$) has been a recurrent problem in the
development of physical cosmology. In this letter we have
discussed a new determination of the Hubble constant based on the
SZE/X-ray distance technique for a sample of 24 clusters as
compiled by De Filippis {\it{et al.}} (2005). The degenerescency
on the $\Omega_{m}$ parameter was broken using the baryon acoustic
oscillation signature from the SDSS catalog. The Hubble constant
was constrained to be $h = 0.74^{+0.04}_{-0.035}$ and
$^{+0.08}_{-0.07}$ for $1\sigma$ and $2\sigma$, respectively.
These limits were derived assuming elliptical $\beta$-model and a
flat $\Lambda$CDM scenario.

As we have seen, the baryon acoustic signature is an interesting
tool for constraining directly the mass density parameter,
$\Omega_m$, and, indirectly, it also improves the Hubble constant
limits acquired from other cosmological techniques (like the
SZE/X-ray cluster distance). Our Hubble constant estimation using
the joint analysis SZE/X-ray + BAO is largely consistent with some
recent cosmological observations, like the third year of the WMAP
and the HST Key Project. Implicitly, such an agreement suggests
that the elliptical morphology describing the cluster sample and
the associated isothermal $\beta$-model is quite realistic. It
also reinforces the interest to the observational search of galaxy
clusters in the near future, when more and larger samples, smaller
statistic and systematic uncertainties will improve the limits on
the present value of the Hubble parameter.

\section*{Acknowledgments}
The authors are grateful to J. G. Bartlett and R. C. Santos for
helpful discussions. JVC is supported by FAPESP No. 2005/02809-5.
JASL is partially supported by CNPq and FAPESP No. 04/13668-0.

{}

\clearpage
\begin{deluxetable}{lccc}
\tablewidth{0pt} \tabletypesize{\footnotesize} \tablecaption{Limits
to $h$ from galaxy clusters ($\Lambda$CDM)\label{Tab1}}
% ***********************************************
\tablehead{
% ***********************************************
{SZ/X-ray method} \\ \hline \colhead{Reference (data)} &
 \colhead{$\Omega_m$} &
 \colhead{$h$ ($1\sigma$)} &
 \colhead{$\chi^2$} \\
}
% ***********************************************
\startdata
% ***********************************************
Mason {\it{et al.}} 2001 (7)&$0.3$& $0.66^{+0.14}_{-0.11}$&$\simeq
2$  \\ Reese {\it{et al.}}
2002(18)&$0.3$&$0.60^{+0.04}_{-0.04}$&$16.5$ \\ Reese 2004
(41)&$0.3$&$0.61^{+0.03}_{-0.03}$&-- \\ Jones {\it{et al.}} 2005
(5)&$0.3$&$0.66^{+0.11}_{-0.10}$&-- \\ Bonamente {\it{et al.}}
2006 (38)&$0.3$&$0.77^{+0.04}_{-0.03}$&$31.6$ \\ {\bf Present work
(24)}&{\boldmath{$0.15^{+0.57}_{-0.15}$}}&{\boldmath{$0.75^{+0.07}_{-0.07}$}}&{\boldmath{$24.4$}}
\\
{\bf{Present work
(24)+BAO}}&{\boldmath{$0.27^{+0.04}_{-0.03}$}}&{\boldmath{$0.74^{+0.04}_{-0.03}$}}&{\boldmath{
$24.5$}}
\enddata
\end{deluxetable}
\clearpage

\end{document}